\documentstyle[12pt,aaspp]{article}
\begin{document}
\title{A LINE OF SIGHT INTEGRATION APPROACH TO COSMIC MICROWAVE BACKGROUND
ANISOTROPIES}

\author{Uro\v s Seljak}
\affil{ Harvard Smithsonian Center for Astrophysics, Cambridge, MA 02138 USA}
\affil{useljak@cfa.harvard.edu}
\author{Matias Zaldarriaga}
\affil{Department of Physics, MIT, Cambridge, MA 02139 USA}
\affil{matiasz@arcturus.mit.edu}
\def\bi#1{\hbox{\boldmath{$#1$}}}
\begin{abstract}
We present a new method for calculating linear cosmic microwave background
(CMB) anisotropy spectra  
based on integration over sources along the photon past 
light cone. In this approach the temperature anisotropy is written as
a time integral over the product of
a geometrical term and a source term. The geometrical term 
is given by radial 
eigenfunctions 
which do not depend on the particular cosmological model. The
source term can be expressed
in terms of photon, baryon and metric perturbations,
all of which can be calculated using a small number of differential 
equations. This split clearly separates between the dynamical 
and geometrical effects on the CMB anisotropies. More importantly, it
allows to significantly reduce the 
computational time compared to standard methods. This is achieved
because the source term, which depends on the model and is
generally the most time consuming part of calculation, is a slowly 
varying function of wavelength and
needs to be evaluated only in a small number of 
points. 
The geometrical term, which 
oscillates much more rapidly than the source 
term, does not depend on the particular 
model and can be precomputed in advance. 
Standard methods that do not 
separate the two terms and require a much higher number of evaluations.
The new method 
leads to about two orders of magnitude reduction in CPU 
time when compared to standard methods and typically 
requires a few minutes on a workstation for 
a single model. The method should be especially useful for 
accurate determinations of cosmological parameters from
CMB anisotropy and polarization
measurements that will become possible with the next 
generation of experiments.
A programm implementing this method can be obtained from
the authors.
\end{abstract}

\newpage
\section{Introduction}

The field of cosmic microwave background (CMB) anisotropies has seen 
a rapid development since its first detection by the COBE satellite
only a few years ago. There are now several reported experimental 
results that are detecting anisotropies
on degree angular scales (see \cite{Scott95} and
\cite{Bond96} for a recent review), 
which together with a few upper limits on smaller 
angular scales already give interesting limits on 
cosmological models. 
With 
the development of the new generation of experiments now being 
proposed one hopes to accurately map the CMB sky 
from arcminute scales to several degree scales. 
The amount of data thus provided would 
allow for an unprecedented accuracy in the determination of cosmological 
parameters. As theoretical modelling shows 
(\cite{Bond94,Hu95c,Jungman95,Seljak94}), 
CMB anisotropies are 
sensitive to most of the cosmological parameters and have a distinctive
advantage over other cosmological observations in that they probe the universe
in the linear regime. This avoids the complications caused by physical 
processes in the nonlinear regime and allows to use powerful statistical
techniques to search over the parameter space for the best cosmological 
model (see e.g. \cite{Jungman95}). 
A large stumbling block in this program at present is the speed of 
theoretical model calculations, which are still too slow to allow for
a rapid search over the parameter space. This limitation was partially 
removed by the development of appoximation methods (\cite{Hu95a},b; 
\cite{Seljak94}), 
which can give fast predictions of CMB anisotropy with a 10\%
accuracy. However, 
these approximations are not accurate enough to exploit the complete
amount of information that will be present in the future CMB observations.
This is especially true for some of the more extreme cosmological models, 
where simple approximations break down and lead to systematic 
inaccuracies in the results.
Obviously, it would be useful to have a fast method that would not 
be based on any approximations and would lead to accurate results for 
any cosmological model. The purpose of this paper is to present a new
method of CMB calculation that satisfies these requirements.

Theoretical calculations of the CMB anisotropies are based on 
linear theory of cosmological perturbations, developed first by 
Lifshitz (1946) and applied to the CMB anisotropies by 
Peebles and Yu (1970). In this early calculation only photons and 
baryons were included, but later workers extended the calculations to 
include dark matter (Bond \& Efstathiou 1984, 1987; Vittorio \& Silk 1984),
curvature (Wilson \& Silk 1981; Sugiyama \& Gouda 1992; 
White \& Scott 1995), 
gravity waves or tensor modes (Crittenden 
et al. 1993) and massive neutrinos (Bond \& Szalay 1983; Ma \& Bertschinger 
1995; Dodelson, Gates \& Stebbins 1995). 
Most of these and more recent calculations (e.g. 
Holtzmann 1989; Stompor 1994; Sugiyama 1995) solve for each Fourier mode of
temperature anisotropy $\Delta_T(\vec k)$ by 
expanding it in Legendre series up to some desired $l_{max}$ and then 
numerically evolve this system of equations in time from the radiation 
dominated epoch until today. Typically this means evolving a system 
of several thousand coupled differential equations in time, a slow 
process even for the present day computers. In addition, because 
each multipole moment is a rapidly oscillating function one has to 
densely sample it in values of 
$k$ with typical number of evaluations of the order of $l_{max}$.
Even the fastest codes at present 
require several hours of 
CPU time for each theoretical model
(Sugiyama 1995), 
while some numerically more accurate ones (e.g. \cite{BB95})
require more like tens or hundreds of hours.

In this paper we explore a different approach to compute CMB anisotropies
based on integration of the sources over the photon
past light cone. The method is a generalization 
of approximate method developed by one of the authors (Seljak 1994).
It differs from it in that it is exact, in the sense that
it can achieve arbitrary precision within the limits of linear 
perturbation theory. By rewriting the system of equations in the
integral form one can separate between the geometrical and dynamical
contributions to the anisotropies. The former do not depend on the 
model and need to be computed only once, while the latter contain all
the information on the particular model and can be computed with a 
small system of equations. Solving for CMB anisotropies using this
integral form greatly reduces the required computational time. 
The outline of the paper is as follows:
in \S 2 we present the basic system of equations that needs to be solved 
both in the standard and in the integral method.
In \S 3 we present in some detail a practical 
implementation of the integral method, 
highlighting the computational 
differences between it and the standard Boltzmann method.
We conclude in \S 4 by discussing 
possible applications where the new method can be particularly useful.

\section{Method}

In this section we first present the standard system of equations 
that needs to be solved for temperature
anisotropies, which is based on solving Boltzmann equation using 
Legendre expansion of photon distribution function. 
This part follows closely 
the existing literature (e.g. Ma \& Bertschinger 1995, Bond 1996
and references therein) 
and only the final results are given. We do
not discuss the technical details of the standard Boltzmann  
method, except where our approach differs significantly from it. 
In the second
part of the section we present the integral solution of photon 
distribution, 
which is the basis of our method.  
In this paper we restrict the analysis to a spatially flat universe. 

\subsection{Boltzmann, Einstein and Fluid Equations}

The temperature anisotropy at position $\vec{x}$ 
in the direction $\vec n$ 
is 
denoted with $\Delta_T(\vec x, \vec n)$. 
In principle it depends both on the direction and 
on the frequency, but because spectral distortions are only introduced
at the second order the frequency dependence can in the lowest order
be integrated out. Anisotropy $\Delta_T(\vec x,\vec n)$ can
be expanded in terms of Fourier modes $\Delta_T(\vec k,\vec n)$,
which in linear perturbation theory evolve independently of 
one another. Assuming perturbations are axially-symmetric 
around $\vec k$, we may further Legendre
expand the anisotropy in the angle 
$\mu=\vec k \cdot \vec n/k$,
\begin{equation}
\Delta_T(\vec k, \vec n)=\sum_l (2l+1)(-i)^l\Delta_{Tl} P_l(\mu),
\label{delta}
\end{equation}
where $P_l(\mu)$ is the Legendre polynomial of order $l$ and $\Delta_{Tl}$ 
is the associated multipole moment.
A similar decomposition also applies to 
the amplitude of polarization anisotropy $\Delta_P(\vec{k},\vec{n})$ 
(\cite{BE84}, Critenden et al. 1993,
Kosowsky 1995, Zaldarriaga \& Harari 1995). 

Evolution of the temperature anisotropy is governed by the Boltzmann
equation (\cite{Peebles70}, \cite{Wilson81}, 
\cite{BE84}). Its collisionless part is given by the time component
of the geodesic equation, which depends on the metric. Here we will 
use the metric 
in the longitudinal gauge (\cite{bard80}; Bertschinger 1996), 
which is similar to the gauge-invariant formalism (Kodama \& Sasaki 1984) 
and gives expressions that are most similar to their  
Newtonian counterparts.
The choice of gauge is purely a matter of convenience
and in some cases (e.g isocurvature models)
other gauge choices such as synchronous gauge 
are computationally advantageous over the longitudinal 
gauge (see e.g. \cite{BB95}, \cite{Bond96}). In the longitudinal gauge
the perturbations are specified with two scalar potentials $\phi$ and
$\psi$ and a gauge-invariant 
tensor perturbation $h$ (we will ignore vector perturbations in this paper,
as they most likely have a negligible contribution to CMB anisotropy).
The corresponding temperature and polarization 
anisotropies are denoted as $\Delta^{(S)}_T$, $\Delta^{(S)}_P$
for scalar and $\Delta^{(T)}_T$, $\Delta^{(T)}_P$
for tensor components. In linear perturbation theory the scalar 
and tensor perturbations  
evolve independently and the total power is given by the sum of the 
two contributions.

The collisional part of the photon Boltzmann equation is determined by the
Thomson scattering term. After angular and momentum 
integration the Boltzmann evolution equations 
for scalar perturbations can be written as (\cite{BE87}),
\begin{eqnarray} 
\dot\Delta_T^{(S)} +ik\mu \Delta_T^{(S)} 
&=&\dot\phi-ik\mu \psi+\dot\kappa\{-\Delta_T^{(S)} +
\Delta_{T0}^{(S)} +i\mu v_b +{1\over 2}P_2(\mu)\Pi
\} \nonumber \\   
\dot\Delta_P^{(S)} +ik\mu \Delta_P^{(S)} &=& \dot\kappa \{-\Delta_P^{(S)} +
{1\over 2} [1-P_2(\mu)] \Pi\} \nonumber \\
\Pi&=&\Delta_{T2}^{(S)}
+\Delta_{P2}^{(S)}+
\Delta_{P0}^{(S)}.
\label{Boltzmann}
\end{eqnarray}
Here the derivatives are taken with respect to the conformal time $\tau$ 
and $v_b$ is the velocity of baryons.
Differential optical depth for Thomson scattering is denoted as 
$\dot{\kappa}=an_ex_e\sigma_T$, where $a(\tau)$ 
is the expansion factor normalized
to unity today, $n_e$ is the electron density, $x_e$ is the ionization 
fraction and $\sigma_T$ is the Thomson cross section. The total optical 
depth at time $\tau$ is obtained by integrating $\dot{\kappa}$,
$\kappa(\tau)=\int_\tau^{\tau_0}\dot{\kappa}(\tau) d\tau$.
A useful variable is the visibility function $g(\tau)=\dot{\kappa}
{\rm exp}(-\kappa)$. Its peak  
defines the epoch of recombination, when the  
dominant contribution to the CMB anisotropies arises.

Expanding the 
temperature anisotropy in multipole moments 
one finds the following hierarchy of coupled 
differential equations (\cite{Wilson81}, \cite{BE84}, \cite{Ma95}),
\begin{eqnarray}
     \dot{\Delta}_{T0}^{(S)}
     &=& -k\Delta_{T1}^{(S)}        +\dot{\phi}
     \,, \nonumber\\
\dot{\Delta}_{T1}^{(S)} &=&
{k \over 3}\left[ \Delta_{T0}^{(S)}-
2\Delta_{T2}^{(S)}
+\psi\right] + \dot{\kappa} ({v_b \over 3}-\Delta_{T1}^{(S)})\,,\nonumber\\
\dot{\Delta}_{T2}^{(S)} &=&{k \over 5}\left[2\Delta_{T1}^{(S)}-3
\Delta_{T3}^{(S)}\right] +
\dot{\kappa} \left[{\Pi \over 10}-\Delta_{T2}^{(S)}\right]\,, \nonumber\\
\dot{\Delta}_{Tl}^{(S)}&=&{k \over
2l+1}\left[l\Delta_{T(l-1)}^{(S)}-(l+1)
\Delta_{T(l+1)}^{(S)}\right]
-\dot{\kappa}\Delta_{Tl}^{(S)} \,, l>2 \nonumber \\
\dot{\Delta}_{Pl}^{(S)}&=&{k \over
2l+1}\left[l\Delta_{P(l-1)}^{(S)}-(l+1)
\Delta_{P(l+1)}^{(S)}\right]
+\dot{\kappa}\left[-\Delta_{Pl}^{(S)}+{1 \over 2}\Pi\left(
\delta_{l0}+{\delta_{l2} \over 5}\right)\right],
\label{photon}
\end{eqnarray}
where $\delta_{ij}$ is the Kronecker symbol.
A similar system of equations without the Thomson scattering terms  
and polarization
also applies for massless neutrinos.
For the
massive neutrinos the system of equations is more complicated,
because the momentum dependence cannot be integrated out of the
expressions (see e.g. \cite{Ma95}).

Baryons and cold dark matter can be approximated as fluids and 
their evolution can be obtained 
from the local conservation of energy-momentum tensor. This gives the 
equations for cold dark matter density $\delta_c$ and its 
velocity $v_c$,
\begin{equation}
\label{cdm2}
        \dot{\delta_c} = -kv_c + 3\dot{\phi}\,, \quad
        \dot{v}_c = - {\dot{a}\over a}\,v_c+k\psi \,.
\end{equation}
For baryons one has additional terms in the Euler's equation
caused by Thomson scattering and pressure,
\begin{eqnarray}
\label{baryon2}
\dot{\delta}_b &=& -kv_b + 3\dot{\phi} \,, \nonumber\\
\dot{v}_b &=& -{\dot{a}\over a}v_b
+ c_s^2 k\delta_b
+ {4\bar\rho_\gamma \over 3\bar\rho_b}
 \dot{\kappa}(3\Delta_{T1}^{(S)}-v_b) + k\psi\,.
 \label{cdmb}
 \end{eqnarray}
Here $c_s$ is the baryon sound speed and $\bar\rho_\gamma$, $\bar\rho_b$
are the mean photon and baryon densities, respectively.

Finally, the evolution of scalar metric perturbations is given by Einstein
equations, which couple the sources and the metric perturbations. Only 
two equations are needed to specify the evolution. Here we choose them to
be the energy and momentum constraint equations,
\begin{eqnarray}
k^2 \phi+3{\dot{a}\over a}\left(\dot{\phi}+{\dot{a}\over a}\psi\right)
=-4\pi Ga^2\delta \rho \nonumber \\
k^2\left(\dot{\phi}+{\dot{a}\over a}\psi\right)=4\pi Ga^2 \delta f,
\label{einstein}
\end{eqnarray}
where $\delta \rho$ and $\delta f$ are the total 
density and momentum density perturbations, respectively. They are 
obtained by summing over the contributions from all species, $\delta \rho
= \sum_i \delta \rho_i$, $\delta f=\sum_i \delta f_i$, $\delta \rho_i=
\bar{\rho_i}\delta_i$ and $\delta f_i=(\bar{\rho_i}+\bar{p_i})v_i$, where 
$\bar{\rho_i}$ and $\bar{p_i}$ are the mean density and pressure of the $i$-th
species.

For tensor perturbations the 
Boltzmann equation is given by (Crittenden et al. 1993),
\begin{eqnarray}
\dot{\Delta}_{T }^{(T)} + i k \mu
 \Delta_{T }^{(T)}&=& - {\dot h}- 
\dot{\kappa} (\Delta_{T}^{(T)} -
\Psi) \ , \nonumber \\
\dot{\Delta}_{P}^{(T)} + i k \mu 
\Delta_{P}^{(T)} &=& - 
\dot{\kappa}( \Delta_{P}^{(T)} + 
\Psi) \ , \nonumber \\
\Psi \equiv  \Biggl\lbrack
{1\over10}\Delta_{T0}^{(T)}
&+&{1\over 35}
\Delta_{T2}^{(T)}+ {1\over210}
\Delta_{T4}^{(T)}
 -{3\over 5}\Delta_{P0}^{(T)}
+{6\over 35}\Delta_{P2}^{(T)}
-{1\over 210}
\Delta_{P4}^{(T)} \Biggr\rbrack . 
\label{tensorl}
\end{eqnarray}
The only external source is that of the tensor metric perturbation 
which evolves according to the Einstein equations as
\begin{equation}
\ddot{h}+2 {\dot{a}\over a}\dot{h}+k^2h=0.
\label{tensoreinstein}
\end{equation}
We ignored the source term on the right-hand side of equation above 
(caused by neutrino and photon anisotropic stress), 
as it is always negligible compared to the terms on the left-hand side.

To obtain the temperature anisotropy for a given mode $\vec k$ one has to
start at early time in the radiation dominated epoch with initial 
conditions of the appropriate type (e.g. isentropic or isocurvature) 
and evolve the system of equations until the present. The anisotropy
spectrum is then obtained by integrating over the initial 
power spectrum of the metric perturbation $P_\psi(k)$,
\begin{equation}
C_l^{(S)}=(4\pi)^2\int k^2dk P_\psi(k)|\Delta^{(S)}_{Tl}(k,\tau=\tau_0)|^2.
\label{cl}
\end{equation}
Analogous expression holds for the polarization spectrum and for the
tensor spectrum (where the initial power spectrum 
$P_\psi(k)$ has to be replaced by the initial tensor power spectrum
$P_h(k)$). 

The spectrum $C_l$ is related to the angular correlation function,
\begin{equation}
C(\theta)=\langle \Delta(\vec n_1)\Delta(\vec n_2)\rangle_{\vec n_1
\cdot
\vec n_2=\cos \theta}={1 \over 4\pi}\sum_{l=0}^\infty(2l+1)C_lP_l(\cos \theta).
\label{ctheta}
\end{equation}
To test a model on a given angular scale
$\theta$ one has to solve for $\Delta_{Tl}$ up to 
$l \approx 1/\theta$. 
If one is interested in small angular scales 
this leads to a large system of differential 
equations to be evolved in time and
the computational time becomes long. For a typical spectrum with 
$l_{\rm max} \sim 1000$ one has to evolve a system of 3000 differential 
equations (for photon and neutrino anisotropy and photon
polarization)
until the present epoch. Moreover, the solutions are rapidly oscillating
functions of time, so the integration has to proceed in small time
increments to achieve the required accuracy on the final values. 

\subsection{Integral solution}

Instead of solving the coupled system of differential equations (\ref{photon})
one may formally integrate equations \ref{Boltzmann} 
along the photon past light cone to obtain (e.g. 
Zaldarriaga \& Harari 1995),
\begin{eqnarray}
\Delta_T^{(S)} &=&\int_0^{\tau_0}d\tau e^{i k \mu (\tau -\tau_0)}
e^{-\kappa} 
\{ \dot\kappa e^{-\kappa} [\Delta_{T0}+i\mu v_b + {1\over 2} P_2(\mu)
\Pi]+\dot\phi-ik\mu\psi\} \nonumber \\
\Delta_P^{(S)} &=& -{1\over 2}\int_0^{\tau_0} d\tau e^{i k \mu (\tau -\tau_0)}
e^{-\kappa} \dot\kappa   [1-P_2(\mu)]
\Pi .
\label{formal}
\end{eqnarray}
Expressions above can be further modified by eliminating the angle $\mu$ in 
the integrand through the integration by parts. 
The boundary terms can be dropped, because they
vanish as $\tau \rightarrow 0$ and are unobservable for $\tau=\tau_0$
(i.e. only the monopole term is affected).
This way each time a given term is multiplied by a $\mu$, it
is replaced by its time derivative. This manipulation leads
to the following expression,
\begin{eqnarray}
\Delta^{(S)}_{T,P} &=&\int_0^{\tau_0}d\tau e^{i k \mu (\tau -\tau_0)}
S^{(S)}_{T,P}(k,\tau) 
\nonumber \\
S^{(S)}_T(k,\tau)&=&g\left(\Delta_{T0}+\psi-{\dot{v_b} \over k}-{\Pi \over 4}
-{3\ddot{\Pi}\over 4k^2}\right)\nonumber \\
&+& e^{-\kappa}(\dot{\phi}+\dot{\psi})
-\dot{g}\left({v_b \over k}+{3\dot{\Pi}\over 4k^2}\right)-{3 \ddot{g}\Pi \over
4k^2} \nonumber \\
S^{(S)}_P(k,\tau)&=&-{3 \over 4 k^2}\left(g\{k^2\Pi+\ddot{\Pi}\}+2
\dot{g}\dot{\Pi}+\ddot{g}\Pi\right).
\label{source}
\end{eqnarray}
Some of the terms in the source function $S^{(S)}_T(\tau)$ are 
easily recognizable. The first two contributions in the first term
are the 
intrinsic anisotropy and gravitational potential 
contributions from the last-scattering surface, while 
the third contribution
is part of the velocity term, the other part being the
$k^{-1}\dot{g}v_b$ term in the second row. 
These terms 
make a dominant contribution to the 
anisotropy in the standard recombination models. 
The first term in 
the second row, 
$e^{-\kappa}(\dot{\phi}+\dot{\psi})$, is the so-called integrated
Sachs-Wolfe term and is important after recombination.
It is especially important if matter-radiation equality occurs 
close to the recombination or in $\Omega_{\rm matter}\ne 1$ models. In both
cases gravitational 
potential decays with time, which leads to an enhancement of 
anisotropies on large angular scales.
Finally we have the terms 
caused by photon polarization and anisotropic
Thomson scattering, which contribute to $\Pi$.
These terms affect the 
anisotropy spectra at the 10\% level and are important for accurate
model predictions. Moreover, they are the sources 
for photon polarization.
Equation (\ref{source}) is a generalization of the tight-coupling
and instantaneous recombination approximation 
(Seljak 1994) and reduces to it in the limit where the visibility function
is a delta-function and $\Pi$ can be neglected. 
In that approximation one only needs to evaluate the sources at 
recombination and then free stream them to obtain the anisotropy today. 
In the more general case presented here
one has to perform an additional integration over time, which includes
the contributions arising during and after the recombination.
Moreover, because the
tight-coupling approximation is breaking down at the time of recombination,
both polarization and photon anisotropic stress 
are being generated and
$\Pi$ makes a non-negligible contribution to the anisotropy.
For exact calculations one has to use (\ref{source}), which 
properly includes all the terms that are relevant in 
the linear perturbation theory.

To solve for the angular power spectrum one
has to expand the plane wave $e^{i k \mu (\tau -\tau_0)}$ 
in terms of the radial and angular eigenfunctions (spherical 
Bessel functions and Legendre polynomials, respectively), 
perform the ensemble average 
\footnote[1]{In performing 
ensemble average we assume that only the amplitude and 
not the phase of a 
given mode evolves in time. While this is 
valid in linear theory for most models of structure formation,
it may not be correct in some versions
of models with topological defects (Albrecht et al. 1995).} 
and integrate over the angular variable 
$\mu$. This leads (\ref{cl}), where 
the multipole moment at present time $\Delta^{(S)}_{(T,P)l}
(k,\tau=\tau_0)$
is given by the following expression,
\begin{equation}
\Delta^{(S)}_{(T,P)l}(k,\tau=\tau_0)=
\int_0^{\tau_0}S^{(S)}_{T,P}(k,\tau)j_l[k(\tau_0-\tau)]d\tau,
\label{finalscalar}
\end{equation}
where $j_l(x)$ is the spherical Bessel function. 
Note that
while angular eigenfunctions integrated out after angular averaging, 
radial eigenfunctions
remained and enter in (\ref{finalscalar}).
The main advantage of (\ref{finalscalar})
is that it decomposes the anisotropy into a source term 
$S^{(S)}_{T,P}$, which does 
not depend on the multipole moment $l$ and a geometrical term $j_l$, which 
does not depend on the particular cosmological model. 
The latter thus only needs to be computed once and can be stored for 
subsequent calculations. 
The source term 
is the same for all multipole moments and only depends on a small number
of contributors in (\ref{source})
(gravitational potentials, baryon velocity and photon moments
up to $l=4$). By specifying the source term as a function of time one can 
compute the corresponding spectrum of anisotropies. 
Equation (\ref{finalscalar})
is formally an integral system of equations, because 
$l<4$ moments appear on both sides of equations. To solve for these
moments it is best to use the equations in their 
differential form (\ref{photon}),
instead of the integral form above.
Once the moments that enter into the source function are computed
one can solve for the higher moments 
by performing the integration in (\ref{finalscalar}) 
(see section 3 for more details).

The solution for the tensor modes can similarly be written as an integral 
over the source term and the tensor spherical eigenfunctions $\chi^l_k$. 
The latter are 
related to the spherical Bessel functions (\cite{abbott86}),
\begin{equation}
\chi^l_k(\tau)=\sqrt{{(l+2)! \over 2(l-2)!}}
{j_l(k\tau)
\over (k\tau)^2}.
\label{eigentensor}
\end{equation}
This gives
\begin{equation}
\Delta^{(T)}_{(T,P)l}=\int_0^{\tau_0}d\tau S_{T,P}^{(T)}(k,\tau)
\chi^l_k(\tau_0-\tau),
\label{finaltensor}
\end{equation}
where from equation \ref{tensorl} follows
\begin{equation}
S_T^{(T)}=-\dot{h}e^{-\kappa}+g\Psi \,\,\,\,\,\,\,\,\,\,\,\,\,\,\,\,
 S_P^{(T)}=-g\Psi.
\label{sourcet}
\end{equation}
Equations (\ref{source}-\ref{sourcet}) 
are the main equations of this paper and form the 
basis of the line of sight integration method
of computing CMB anisotropies. In the next section we will
discuss in more detail the computational advantages of this 
formulation of Boltzmann equation and its implementation.

\section{Calculational Techniques}

In the previous section we presented the expressions needed 
for the implementation of the line of sight integration
method. 
As shown in  (\ref{finalscalar}) and (\ref{finaltensor})
one needs to integrate over
time the source term at time $\tau$ 
multiplied with the sperical Bessel function
evaluated at $k(\tau_0-\tau)$. The latter does not depend
on the model and can be precomputed in advance. 
Fast algorithms exist which can compute spherical Bessel functions
on a 
grid in $k$ and $l$ in short amount of time (e.g. Press et al. 1993). The
grid is then stored on a disk and used for all the subsequent calculations.
This leaves us with the
task of accurately calculating the source term,
which determines the CMB spectrum for a given model.
Below we discuss some of the calculational techniques needed for
the implementation of the method. 
We especially highlight the differences
between this approach and the standard Boltzmann
integration approach. Our goal is to develop a method which is
accurate to 1\% in $C_l$ up to $l \sim 1000$
over the whole range of cosmological 
parameters of interest. These include models with varying amount of
dark matter, baryonic matter, Hubble constant, vacuum energy, 
neutrino mass, shape of initial spectrum of perturbations, reionization
and tensor modes. The choice of accuracy is based on estimates of
observational accuracies that will be achiavable in the next generation 
of experiments and also on the theoretical limitations of model 
predictions (e.g. cosmic variance, second order effects etc.).
Most of the figures where we discuss the choice
of parameters 
are calculated for the standard CDM model. 
This model is a reasonable choice in the sense that it is a model which 
exhibits most of the physical effects in realistic models, including 
acoustic oscillations, early-time integrated Sachs-Wolfe effect and 
Silk damping. One has to be careful however not to tune the parameters
based on a single model. We compared our results with results from other 
groups (Bode \& Bertschinger 1995; Sugiyama 1995) for a number of 
different models. We find a better than 1\% agreement 
with these calculations over most of the parameter space of models.
The computational parameters we recommend below are based on this more detailed
comparison and are typically more stringent than 
what one would find based on the comparison with the standard CDM model only. 

\subsection{Number of coupled differential equations}

\begin{figure}[t]
\vspace*{8.3 cm}
\caption{CMB spectra produced by 
varying the number of evolved photon multipole moments, together
with the  
relative error (in \%)
compared to the exact case. While using $l_\gamma=5$ produces
up to 2\% error, using $l_\gamma=7$ gives results almost identical to the 
exact case.}
\includegraphics{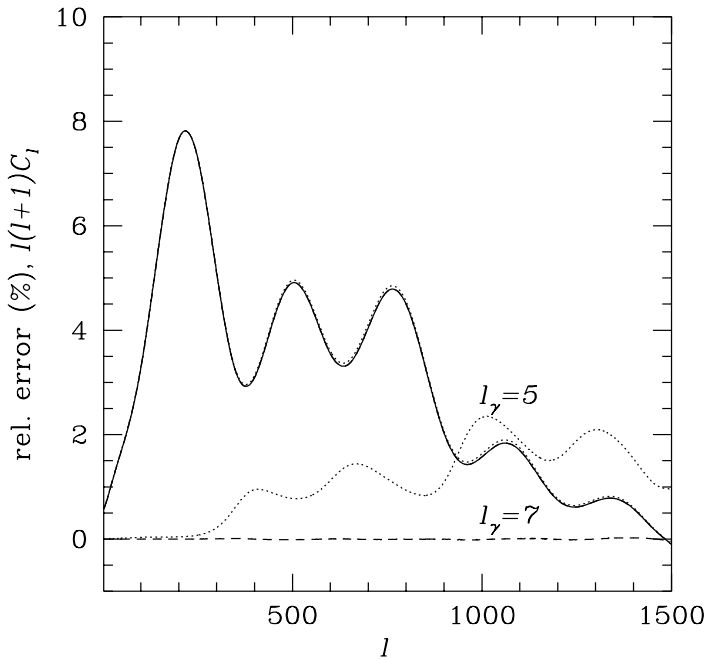}

\label{fig1}
\end{figure}

In the standard Boltzmann method the 
photon distribution function is expanded to a high $l_{\rm max}$ 
(\ref{photon}) and typically one has to 
solve a coupled system of several thousand differential equations.
In the integral method one
evaluates the source terms $S(k,\tau)$ as a function 
of time (\ref{source}), (\ref{sourcet}) and
one only requires the knowledge of photon multipole 
moments up to $l=4$, 
plus the metric perturbations and baryon velocity. 
This greatly reduces 
the number of coupled differential equations that are needed to be solved. 
For an accurate evaluation of the lowest multipoles in the integral
method one has to extend
the hierarchy somewhat beyond $l=4$, because the
lower multipole moments are coupled to the higher multipoles
(\ref{photon}).  
Because power is only being 
transferred from lower to higher $l$  
it suffices to keep a few 
moments to achieve a high numerical accuracy of $l<5$ moments.
One has to be careful however to avoid 
unwanted 
reflections of the power being transferred from low $l$ to high $l$, 
which occur for example if a simple cut-off in the hierarchy is imposed.
This can be achieved by modifying 
the boundary condition for the last term in the 
hierarchy 
using the free streaming approximation
(\cite{Ma95}, \cite{Hu95}). In the absence of scattering 
(the so-called free streaming regime),
the recurrence relation among the 
photon multipoles in equation (\ref{photon}) becomes the
generator of spherical Bessel functions. 
One can therefore use a different recurrence 
relation among the spherical Bessel functions 
to approximate the last term in the hierarchy without reference to the 
higher terms.
The same approximation can also be used for polarization and
neutrino hierarchies. This type of
closure scheme works extremely well and only a few multipoles
beyond $l=4$ are needed for an accurate calculation of the source 
term. This is shown in 
figure \ref{fig1}, where a relative error in the spectrum is plotted
for several choices of maximal number of photon multipoles.
We choose to end the
photon hierarchy (both anisotropy and polarization) 
at $l_\gamma=8$ and massless neutrino at $l_\nu=7$, which 
results in an error lower than $0.1\%$ compared to the exact case. 
Instead of a few thousand coupled differential equations 
we therefore evolve about 35 equations
and the integration time is correspondingly reduced. 

\subsection{Sampling of CMB multipoles}

\begin{figure}[t]
\vspace*{9.3 cm}
\caption{  
Relative error between the exact and interpolated spectrum,  
where every 20th, 50th or 70th multipole is calculated. The 
maximal error for the three approximations is less
than 0.2, 0.4 and 1.2\%, respectively.  
The rms deviation from the exact spectrum is further improved by
finer sampling, because the
interpolated spectra are exact in the sampled points. For the sampling
in every 50th multipole the rms error is 0.1\%.} 
\includegraphics{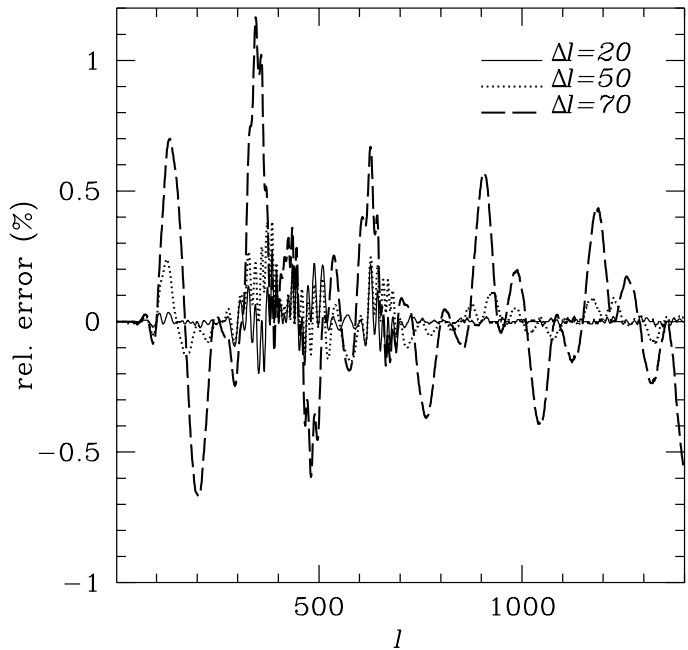}
\label{fig2}
\end{figure}

In the standard Boltzmann integration method one solves for the whole 
photon hierarchy (\ref{photon}) and the 
resultant $\Delta_l$ is automaticaly obtained for each $l$ up to some 
$l_{\rm max}$. 
The CMB spectra are however
very smooth (see figure \ref{fig1}), except for the lowest $l$, where the
discrete nature of the spectrum becomes important. This means that 
the spectrum need not be sampled for each $l$ and instead  
it suffices to sparsely sample
the spectrum in a number of points and interpolate between them.
Figure \ref{fig2} shows the result of such interpolation 
with cubic splines (see e.g. \cite{Press92}) when every 
20th, 50th or 70th 
$l$ is sampled beyond $l=100$ with an increasingly denser sampling
towards small $l$, so that each $l$ is sampled below $l=10$. 
While sampling of every 70th $l$ results in maximal error of 
1\%, sampling in 
every 20th or 50th $l$ gives errors below 0.2 and 0.4\%, respectively. 
We choose to compute every 50th $C_l$ beyond $l=100$ in addition to
15 $l$ modes
below $l=100$, so that a total of
45 $l$ modes are calculated up to $l_{\rm max} =1500$. This gives   
a typical (rms) error of 0.1\%, with excursions of up to 0.4\%. 
The number of integrals in equation (\ref{cl}) is thus 
reduced by 10-50 and the computational 
time needed for the integrals becomes comparable or smaller
than the time
needed to solve for the system of differential equations.

\subsection{Free streaming}

After recombination and in the absence of a time changing gravitational 
potential the source function often becomes negligible. This is 
the so called free streaming regime, where the
photons are freely propagating through the universe. 
Most of the standard Boltzmann codes use a special free streaming algorithm 
to map the anisotropies from a given epoch $\tau_{\rm fs}$ into 
anisotropies today (\cite{BE84}).
In the line of sight integration method
the free streaming regime is only a special case where $S(k,\tau)=0$ after 
some time $\tau_{\rm fs}$.
Thus one can stop the integration at the 
time $\tau_{\rm fs}$ beyond which the sources are not important and there is
no need for a separate algorithm to evolve the anisotropies until today. 
For example, if one
assumes that only the first term in (\ref{source})
is important
one only needs to integrate over 
the source where the visibility function $g$ appreciably differs from 0. In 
the absence of reionization this restricts the time 
integration to a narrow interval during recombination   
around $z \approx 1100$.
Although most of the contributions to the anisotropies come
from this epoch, time dependent
gravitational potential (and, to a smaller extent, other source terms
in equation (\ref{source})) make 
a nonnegligible contribution to the anisotropies
even after recombination. 
As mentioned earlier,
this is especially important
in $\Omega_{\rm matter}\ne 1$ models and in models with low $\Omega_{
\rm matter}h^2$. In the first case 
the gravitational potential is decaying at late times,
while in the second class of models the matter-radiation equality
during which gravitational potential
changes in time is pushed to a lower redshift.
Even in standard CDM model ($\Omega_{\rm matter}=1,h=0.5$)
gravitational potential is still 
significantly changing in time at moderately low redshifts
of $z \sim 100$ (\cite{Hu95}).
Similarly one cannot use free streaming
in the models with late reionization, 
where the visibility
function is nonvanishing at low redshifts. 
We choose to integrate until the present time 
for most of the models, except for
the models with $\Omega_{\rm matter}=1$, 
where we stop the integration at 
$z=10$. In this case the  
computational time is reduced significantly (typically 50\%) 
compared to that of evolving the 
equations until the present time. 

\subsection{Integration over time}

\begin{figure}[t]
\vspace*{9.3 cm}
\caption{ Error in the spectrum caused by insufficient
temporal sampling of the source term. 
Inaccurate sampling of the source during recombination leads to 
numerical errors, which can reach the level of 1\% if the source 
is sampled in only 10 points across the recombination epoch. 
Finer sampling in time gives much smaller errors for this model. 
Comparisons with other models indicate that sampling in 40 
points is needed for accurate integration.} 

\includegraphics{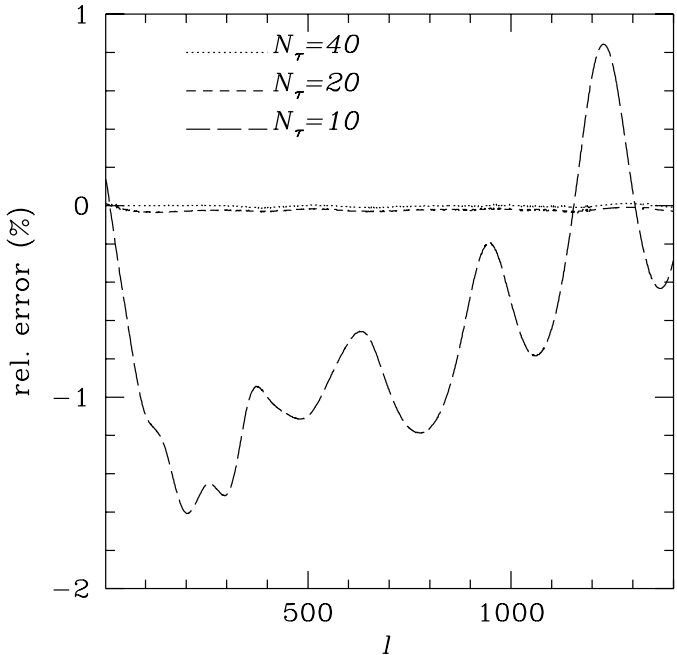}

\label{fig3}
\end{figure}

For each Fourier mode $k$ the source term is integrated
over time $\tau$ (\ref{finalscalar}). 
The sampling in time need not be 
uniform, because the dominant contribution arises from the epoch
of recombination around $z\sim 1100$, the width of which is 
determined by the visibility function $g$ and is rather narrow in 
look-back time for
standard recombination scenarios. During this epoch the sources
acoustically oscillate on a time scale of $c_sk^{-1}$, 
so that the longest wavelength modes are the slowest to vary.
For short wavelengths the rate of sampling should therefore be higher.
Even for long wavelengths the source function
should still be sampled in several
points across the last-scattering surface. This is because the 
terms in (\ref{source}) depend on the derivatives
of the visibility function. 
If the visibility function $g$ is narrow then its derivative will 
also be narrow and will sharply 
change sign at the peak of $g$. Its integration 
will lead to numerical roundoff
errors if not properly sampled, even though  
positive and negative contributions nearly cancel out when integrated
over time and make only a small contribution to the integral. 
Figure \ref{fig3} shows 
the error in integration caused by sampling this epoch with 10, 20
or 40 points. Based on comparison with several models
we choose to sample the recombination epoch
with 40 points, which results in very small ($\sim$ 0.1\%) errors. 
After this 
epoch the main contribution to the anisotropies arises from the 
integrated Sachs-Wolfe term. This is typically a slowly changing
function and it is sufficient to sample 
the entire range in time until the present in 40 points.
The exceptions here are models
with reionization, where the visibility function becomes non-negligible
again and a new last-scattering surface is created. In this case a 
more accurate sampling of the source is also needed at lower redshifts.

\subsection{Integration over wavenumbers}

The main computational cost of standard 
CMB calculations is solving the coupled
system of differential equations.
The number of $k$-modes for which the system is solved is the main 
factor that determines the speed of the method. For results accurate
to $l_{\rm max}$ one has to sample the wavenumbers upto a maximun value
$k_{\rm max}=l_{\rm max}/ \tau_0$.  
In the line of sight integration method 
solving the coupled
system of differential equations still dominates the computational time 
(although for each mode 
the time is significantly shorter than in the 
standard Boltzmann method 
because of a smaller system of equations).
It is therefore instructive to compare the number of $k$ 
evaluations needed in each of the methods to achieve a given
accuracy in the final spectrum.

\begin{figure}[t]

\vspace*{9.3 cm}
\caption{In (a) $\Delta_{T,150}^{(S)}(k)$ is plotted as a function of
wavevector 
$k$. In (b) $\Delta_{T,150}^{(S)}(k)$ is decomposed into the 
source term $S_T^{(S)}$ integrated over time
and the spherical
Bessel function $j_{150}(k\tau_0)$. The high frequency oscillations 
of $\Delta_{T,150}^{(S)}(k)$ are caused by oscillations of 
the spherical Bessel function $j_{150}(k\tau_0)$, whereas the source 
term varies much more slowly. This allows one to reduce the number of
$k$ evaluations in the line of sight integration method, because only 
the source term needs to be sampled.  } 
\includegraphics{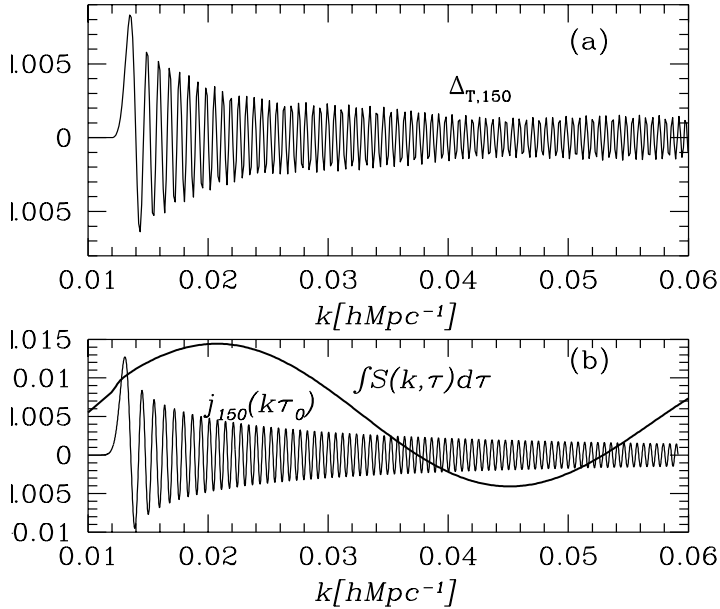}
\label{fig4}
\end{figure}

In the standard Boltzmann method
one solves for $\Delta_{T,l}^{(S)}(k)$ directly, so this quantity 
must be sampled densely enough for accurate integration.
Figure \ref{fig4}a shows $\Delta_{T,l}^{(S)}(k)$
for $l=150$ in a standard CDM model. One can see that it is a rapidly 
oscillating function with a frequency $k \sim \tau_0^{-1}$. 
Each oscillation needs to be sampled in at least a few points to assure 
an accurate integration.
To obtain a smooth CMB spectrum one typically 
requires 6 points over one period, implying 
$2l_{\rm max}$ $k$-mode evaluations. 
This number can be reduced somewhat 
by filtering out the sampling noise in the spectrum (\cite{Hu95}), but even
in this case one requires at least 1-2 points per each period or 
$l_{\rm max}/2$ $k$-mode evaluations. 

To understand the nature of these rapid oscillations in
$\Delta_{T,l}^{(S)}(k)$ we will consider wavelenghts larger than
the width of the last scattering surface.  In this case the
Bessel function in (\ref{finalscalar}) can be pulled out of the integral as
$j_l(k\tau_0)$ because the time at which recombination occurs,
when the dominant contribution to $\Delta_{T,l}^{(S)}(k)$ is
created, is 
much smaller than $\tau_0$ and $k \Delta\tau\ll 1$ ($\Delta\tau$
is the interval of time for which the visibility function differs
appreciably from zero).
So the final $\Delta_{T,l}^{(S)}(k)$ is approximately 
the product of $j_l(k\tau_0)$ and $S_T^{(S)}$ integrated over time,
if the finite width of the last scattering surface and
contributions after recombination can be  ignored.  

Figure \ref{fig4}b shows the 
source term $S_T^{(S)}$ integrated over time
and the Bessel function $j_l(k\tau_0)$. 
It shows that the high frequency
oscillations in $\Delta_{T,l}^{(S)}(k)$ seen in figure \ref{fig4}a
are caused by the oscillation of the spherical
Bessel functions, while the oscillations of the source term have a
much longer period in $k$.
The different periods of the two 
oscillations can be understood
using the tight coupling approximation (\cite{Hu95}, \cite{Seljak94}). 
Prior and during recombination photons are coupled to the baryons and the 
two oscillate together  
with a typical acoustic timescale $\tau_{s}\sim \tau_{\rm rec}/\sqrt{3}
\sim \tau_0/\sqrt{3z_{\rm rec}} \sim \tau_0/50$. The 
frequency of acoustic oscillations $k \sim \tau_{\rm rec}^{-1}$ 
is therefore 50 times higher than
the frequency of oscillations in spherical Bessel functions, which 
oscillate as $\tau_0^{-1}$. 

\begin{figure}[t]
\vspace*{8.3 cm}
\caption{ Error in the spectrum caused by insufficient 
$k$-mode sampling of the source term. 
Sampling the source with 40 points up to 
$k = 2l_{\rm max}$ leads to 1\% errors, while  
with 60 or 80 points the maximal error decreases to 0.2\%.
Comparisons with other models indicate that sampling in 60 
points is sufficient for accurate integration.} 
\includegraphics{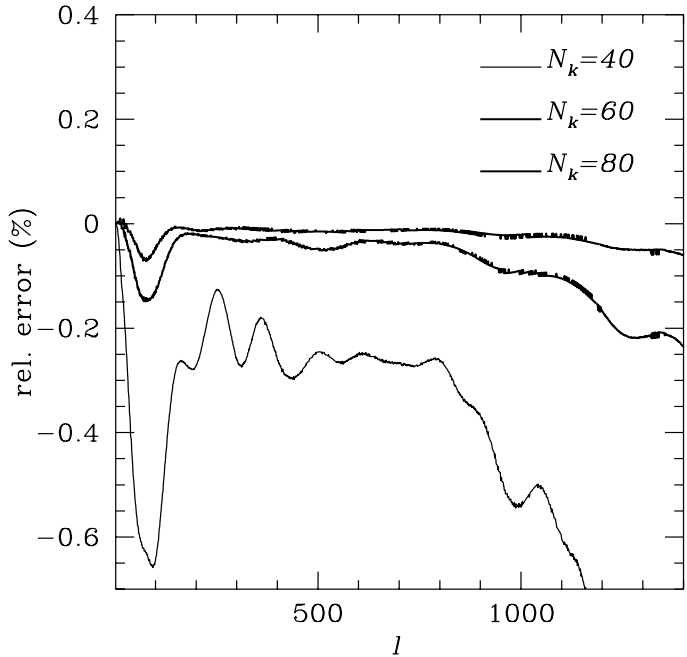}
\label{fig5}
\end{figure}

Because an accurate sampling of the 
source term requires only a few 
points over each acoustic oscillation, the total number of $k$ evaluations
in the integral method 
can be significantly reduced compared to the standard methods. 
Typically a few dozen
evaluations are needed
over the entire range of $k$, compared to about 500 evaluations
in the standard method when a noise filtering technique is 
used and 2000 otherwise 
(for $l_{\rm max} \sim 1000$). Once the source
term is evaluated at these points one can
interpolate it at points with preevaluated 
spherical Bessel functions, which can be much more densely sampled 
at no additional computational cost. 
The end result is the same accuracy as in the standard method,
provided that the source is sampled in sufficient number of points.
Figure \ref{fig5} shows the relative error in the CMB spectrum for the
cases where the 
source term is calculated in 40, 60 and 80 points between 
0 and $k \tau_0 =3000$ (for $l_{\rm max}=1500$). 
While using 40 points results in up to 1\% errors, using 60 points
decreases the maximum error to below 0.2\% for this model. 
In general it suffices to use 
$l_{\rm max}/30$ $k$ modes, which is at least  
an order of magnitude smaller than in 
the standard methods. 
Note that with this method there is no need to 
filter the spectrum to reduce the sampling noise, because  
the latter is mainly caused by insufficient sampling
of the spherical Bessel functions, which
are easy to precompute. The additional operations needed for a higher 
sampling (summation and source interpolation) do not  
significantly affect the overall computational time. Moreover, 
if each $C_l$ is accurately calculated they can be sparsely sampled
and interpolated
(section 3.2), this would not be possible if they had a significant noise
component added to them.

\section{Conclusions}
  
In this paper we presented a new method for accurate calculations of
CMB anisotropy and polarization spectra. 
The method is not based on any approximations and is an
alternative to the standard Boltzmann calculations, which are based 
on solving large numbers of differential equations. The 
approach proposed here uses a hybrid integro-differential 
approach in solving the same system of equations. 
By rewriting the Boltzmann equations in the integral form
the solution for the photon anisotropy 
spectrum can be written as an integral over a
source and a geometrical term. The first is determined by a small number
of contributors to the photon equations of motion and the second is 
given by the radial eigenfunctions, which do not depend on the
particular cosmological model, but only on the geometry of space.

One advantage of the split between geometrical and dynamical 
terms is that
it clarifies their different contributions to
the final spectrum. A good example of this is the temperature anisotropy
in the non-flat universe, which can be be written using a similar 
decomposition, except
that spherical Bessel functions have to be replaced with their 
appropriate generalization (\cite{abbott86}). 
This will be discussed in more detail in a future 
publication, here we simply remark that replacing radial eigenfunctions
in a non-flat space with their flat space counterpart (keeping comoving
angular distance to the LSS unchanged) is only approximate and
does not become exact even in the large
$l$ (small angle) limit. The geometry of the 
universe leaves its signature in the CMB spectra in a rather 
nontrivial way and does not lead only to a simple rescaling of the 
spectrum by $\Omega_{\rm matter}^{-1/2}$ (\cite{Jungman95}).

The main advantage of our line of sight integration method is
its speed and accuracy. 
For a given set of parameters it is two orders of magnitude
faster than the standard Boltzmann 
methods, while preserving the same accuracy.
We compared our results with the results by Sugiyama (1995) and
by Bode \& Bertschinger (1995) and in both cases the agreement was 
better than 1\% up to a very high $l$ for all of the models we
compared to.

The method is useful for fast and accurate normalizations
of density power spectra from CMB measurements, 
which for a given model require the CMB anisotropy spectrum and
matter transfer function, both of which are provided by the output
of the method. 
Speed and accuracy are even more important for accurate determination
of cosmological parameters from CMB measurements. In such applications
one wants to perform a search over a large parameter space,
which typically requires calculating the spectra of a
several thousand models (e.g. \cite{Jungman95}). 
One feasible way to do so is to use approximation methods 
mentioned in the introduction. These 
can be made extremely fast, but at a cost of sacrificing the 
accuracy. While several percent accuracy is sufficient for analyzing
the present day experiments, it will not satisfy the requirements
for the future all-sky surveys of microwave sky. Provided that 
foreground contributions can be succesfully filtered out (see
Tegmark \& Efstathiou 1995 for a recent discussion) one can hope for 
accuracies on the spectrum close to the cosmic variance limit,
which for a broad band averages can indeed reach below 1\% at 
$l>100$. It is at this stage that fast and accurate CMB calculations
such as the one presented in this paper
will become crucial and might enable one 
to determine many cosmological parameters with an unprecedented
accuracy.

\acknowledgements
We would like to thank Ed Bertschinger for encouraging 
this work and providing helpful comments.
This work was partially supported by grant NASA NAG5-2816.

\end{document}